\documentclass[conference]{IEEEtran}
\IEEEoverridecommandlockouts
\usepackage{cite}
\usepackage{amsmath,amssymb,amsfonts}
\usepackage{algorithmic}
\usepackage{graphicx}
\usepackage{epstopdf}
\usepackage{textcomp}
\usepackage{xcolor}
\def\BibTeX{{\rm B\kern-.05em{\sc i\kern-.025em b}\kern-.08em
    T\kern-.1667em\lower.7ex\hbox{E}\kern-.125emX}}
\begin{document}

\title{Delay-Doppler Reversal for OTFS System in Doubly-selective Fading Channels\\
}

\author{\IEEEauthorblockN{ \textbf{Xiangxiang Li$^{1,2}$, Haiyan Wang$^{1,3}$, Yao Ge$^{4}$, Xiaohong Shen$^{1,2}$, Yuanyuan Lei$^{5}$}}
\IEEEauthorblockA{$^{1}$ School of Marine Science and Technology, Northwestern Polytechnical University, Xi’an, Shaanxi, China, 710072 \\
$^{2}$ Key Laboratory of Ocean Acoustics and Sensing (Northwestern Polytechnical University), 
Ministry of \\ Industry and Information Technology, Xi’an, Shaanxi, China, 710072\\
$^{3}$ School of Electronic Information and Artificial Intelligence Shaanxi University of 
 Science and \\ Technology, Xi’an, Shaanxi, 710021, China \\
$^{4}$ Continental-NTU Corporate Lab, Nanyang Technological University, Singapore\\
$^{5}$ Xi’an Electronic Engineering Research Institute, Xi’an 710100, China\\
lixx@mail.nwpu.edu.cn; hywang@sust.edu.cn; yao.ge@ntu.edu.sg; xhshen@nwpu.edu.cn; leiyyiel@163.com\\
}}


\maketitle

\begin{abstract}
	
The recent proposed orthogonal time frequency space (OTFS) modulation shows significant advantages than conventional orthogonal frequency division multiplexing (OFDM) for high mobility wireless communications. However, a challenging problem is the development of efficient receivers for practical OTFS systems with low complexity. In this paper, we propose a novel delay-Doppler reversal (DDR) technology for OTFS system with desired performance and low complexity. We present the DDR technology from a perspective of two-dimensional cascaded channel model, analyze its computational complexity and also analyze its performance gain compared to the direct processing (DP) receiver without DDR. Simulation results demonstrate that our proposed DDR receiver outperforms traditional receivers in doubly-selective fading channels.

\end{abstract}


\section{Introduction}
In recent years, with the widespread development of wireless communication network, the next generation wireless communication technologies are expected to support reliable and high throughput communication for high mobility scenarios such as high-speed trains and vehicle-to-everything (V2X). However, the current widely used orthogonal frequency division multiplexing (OFDM) modulation \cite{b15} will suffer from high inter-carrier interference (ICI) caused by the Doppler spread in the high mobility scenarios, resulting in severe performance loss. In order to tackle the high mobility doubly-selective fading channels and improve system performance, orthogonal time frequency space (OTFS)  modulation was recently proposed in  \cite{b1}. Different from the OFDM, OTFS transmits the information symbols in delay-Doppler (DD) domain instead of time-frequency (TF) domain. In this way, it can transform a rapidly time-varying multipath channel in TF domain to an almost time-invariant channel in DD domain. Such invariant channel model in DD domain can simplify receiver complexity and improve bit error rate (BER) performance compared to OFDM in the high mobility communication systems \cite{b16}.

Although the DD domain channel can be approximated as time-invariant, how to achieve better performance with low complexity is still a challenging problem for wireless receivers. The classic linear equalizers such as linear minimum mean square error (LMMSE) requires a large computational overhead due to matrix inverse \cite{b6}. In order to reduce the equalizer complexity, a lot of nonlinear receiver algorithms were proposed for OTFS by utilizing the channel sparsity in DD domain \cite{b7}, \cite{b14}. The message passing (MP) algorithm with Gaussian approximation was developed for interference cancellation in \cite{b2}, \cite{b13}. However, the receiver performance may degrade significantly due to the loopy effect of the dense factor graph in rich-scattering scenarios. The approximate message passing (AMP) receivers were proposed in  \cite{b3}, \cite{b4}  to further reduce the computational complexity, however, its performance is limited to the large i.i.d. sub-Gaussian channel matrix. The unitary approximate message passing (UAMP) was proposed in \cite{b5} to tackle the influence of sub-Gaussian channel matrix. However, the complexity is still as high as that of LMMSE with large dimension matrix inverse. 


In order to achieve desired performance and alleviate the complexity at the receiver, time reversal (TR) is regarded as a promising solution in the literature, especially for the frequency-selective fading channels \cite{b8}, \cite{b12}. It can exploit the frequency diversity to address the inter-symbol interference (ISI) caused by the multipath channel. However, the traditional TR technology was proposed only for one-dimensional time domain, which can not be directly applied to two-dimensional DD domain OTFS system with doubly-selective fading channels. Recently, a two-dimensional passive time reversal technology was proposed for OTFS in \cite{b9} to achieve spatial, time and frequency focusing with better performance. In this paper, inspired by TR technology, we develop a novel two-dimensional DD reversal (DDR) technology for OTFS system with expected performance and relatively low complexity. Firstly, we present the DDR technology from a perspective of two-dimensional cascaded channel model. Secondly, we compare the signal-to-interference-plus-noise-ratio (SINR) gain between DDR and direct processing (DP) receivers and analyze the computational complexity. Finally, we compare the BER performance of different scenarios and schemes, including the different number of antennas, modulation alphabet, User Equipment (UE) speed and channel uncertainty. Through theoretical analysis and simulation test, we demonstrate that our proposed DDR technology can effectively improve the performance of OTFS system in doubly-selective fading channels and robust to the imperfect channel state information (CSI).

\section{OTFS Transmission Model}

The OTFS can be regarded as a two-dimensional extension of OFDM. It mainly maps the information symbols to DD domain rather than TF domain by adding the pre-processing and post-processing in OFDM systems. Within OTFS framework, the TF signal plane is given by
\begin{align}\nonumber
	\Lambda=\{\left( m\Delta f,nT\right) ,m=0,\cdot\cdot\cdot,M-1,n=0,\cdot\cdot\cdot,N-1\},
\end{align}
with the sampling time and frequency axes at intervals $T(s)$ and $\Delta f=1/T(Hz)$, respectively. The duration and occupyed bandwidth for this TF plane are $ T_f=NT $ and $ B=M\Delta f $, respectively. 

In OTFS system, the $MN$ information symbols $\mathbf{X}_\textnormal{DD} \in \mathbb{C}^{M \times N} $ are firstly placed on DD domain grid
\begin{align}\nonumber
	\Gamma={\left( \frac{l}{M\Delta f},\frac{k}{NT}\right) ,l=0,\cdot\cdot\cdot,M-1,k=0,\cdot\cdot\cdot,N-1},
\end{align}
where $ 1/M\Delta f $ and $ 1/NT $ represent the resolutions of delay and Doppler axes, respectively. Then, the $ MN $ DD domain symbols are mapped into a lattice in TF domain $ \mathbf{X}_\textnormal{TF} \in \mathbb{C}^{M \times N} $ by inverse symplectic finite Fourier transform (ISFFT)
\begin{align}\label{1}
	\mathbf{X}_\textnormal{TF}=\mathbf{F}_M \mathbf{X}_\textnormal{DD} \mathbf{F}^\textnormal{H}_N,
\end{align}
where $\mathbf{F}_M \in \mathbb{C}^{M \times M} $ and  $\mathbf{F}^\textnormal{H}_N \in \mathbb{C}^{N \times N}$ are the normalized $M$-point discrete Fourier trasnform (DFT) matrices and $N$-point inverse discrete Fourier trasnform (IDFT) matrices, respectively. The Heisenberg transform is adopted to the TF signal $\mathbf{X}_\textnormal{TF}$ with a transmit rectangular pulse, thus, the time domain signal $\mathbf{X}_\textnormal{T} \in \mathbb{C}^{M \times N} $ can be expressed as
\begin{align}\label{2}
	\mathbf{X}_\textnormal{T}=\mathbf{F}^\textnormal{H}_M \mathbf{X}_\textnormal{TF}.
\end{align}

Then, the transmitted signal $ \mathbf{x} \in \mathbb{C}^{MN \times 1} $ can be obtained by vectorizing the signal $ \mathbf{X}_\textnormal{T} $, which can be expressed as
 \begin{align}\label{3}
	\mathbf{x}=\mathbf{vec} (\mathbf{{X}}_\textnormal{T}),
\end{align}
where $\mathbf{vec}\{\cdot\}$ is vectorizing operation in column-wise. To overcome the inter-frame interference, we append a cyclic prefix (CP) of length no shorter than the maximal channel delay spread to signal $ \mathbf{x} $.

The DD domain channel $h[l,k]$ with baseband impulse response can be expressed as
\begin{align}\label{4}
	h[l,k]=\sum_{k_\nu=-K/2}^{K/2}\sum_{l_\tau=0}^{L-1}h_{l_\tau,k_\nu}\delta[l-l_\tau]\delta[k-k_\nu],
\end{align}
where the $h_{l_\tau,k_\nu}$, $l_\tau$, $k_\nu$, $L$ and $K/2$ represent the complex gain, delay, Doppler shift, the maximum supportable delay and Doppler shift, respectively. For simplicity, we assume the delay and Doppler shifts as integer multiples of $\frac{1}{M\Delta f}$ and $\frac{1}{NT}$, respectively, i.e., we assume the $ l_\tau $ , $ k_\nu $ are integers. However, fractional delay and Doppler shifts can also be handled using the techniques discussed in \cite{b2} by adding virtual integer taps in the DD domain channel. Hence, the results derived in this paper can be straightforwardly extend to the fractional delay and Doppler shifts.

At the receiver, the baseband received signal $ \mathbf{y} \in \mathbb{C}^{MN \times 1} $ after removing CP in the time domain can be expressed as
\begin{align}\label{5}
	y[c]=\sum_{k=-K/2}^{K/2}\sum_{l=0}^{L-1}h[l,k]x[c-l]_{MN}e^{j2\pi k (c-l)}+v[c], 
\end{align}
where $c=0,1,..., MN-1$ and $ \mathbf{v} \in \mathbb{C}^{MN \times 1}$ is additive white Gaussian noise (AWGN) with zero-mean and variance $\sigma^2$. 

The received signal $ \mathbf{y} $ is then converted into a matrix $ \mathbf{Y}_\textnormal{T} \in \mathbb{C}^{M \times N} $ given by
\begin{align}\label{6}
	\mathbf{Y}_\textnormal{T}=\mathbf{invec}\{\mathbf{y}\},
\end{align}
where $\mathbf{invec}\{\cdot\}$ is an inverse of the $\mathbf{vec}\{\cdot\}$ to transform vector back to a martix. After applying Wigner transform (i.e., the inverse of Heisenberg transform) with a rectangular pulse, the received TF domain signal $\mathbf{Y}_\textnormal{TF} \in \mathbb{C}^{M \times N}$ can be expressed as
\begin{align}\label{7}
	\mathbf{Y}_\textnormal{TF}=\mathbf{F}_M \mathbf{\mathbf{Y}_\textnormal{T}}.
\end{align}

Finally, the signal $\mathbf{Y}_\textnormal{DD} \in \mathbb{C}^{M \times N}$ in DD domain can be obtained by the symplectic finite Fourier transform (SFFT)
\begin{align}\label{8}
	\mathbf{Y}_\textnormal{DD}=\mathbf{F}^\textnormal{H}_M  \mathbf{Y}_\textnormal{TF} \mathbf{F}_N.
\end{align}


\section{Proposed DDR Receiver}

In the literature, TR is regarded as an efficient way to achieve better receiver performance with low complexity \cite{b8}. However, the classic TR is only a one-dimensional reversal in the time domain, which can not be directly applied to the two-dimensional DD domain for OTFS. In order to achieve the focus of paths in DD domain channel, we develop a novel DDR technology through two-dimensional cascaded channel model for OTFS systems in this section.

\begin{figure}[htbp]
	\centerline{\includegraphics[scale=0.48]{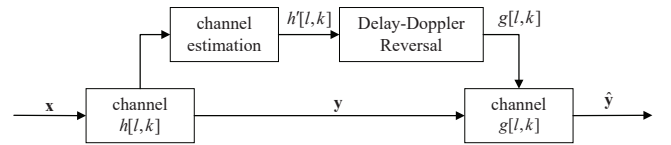}}
	\caption{Basic structure of proposed DDR.}
	\label{figDDR}	
\end{figure}

The basic structure of proposed DDR technology is given in Fig. \ref{figDDR}. Compared to the classic TR, the DDR is consisted of two parts: channel estimation $h'[l,k]$ and DDR channel $g[l,k]$. Here, we assume the result of channel estimation is perfect, i.e. $h'[l,k] = h[l,k]$. In order to improve the SNR at receiver, we design the DD domain channel $g[l,k]$ to match the channel $h[l,k]$, which is given by

\begin{align}\label{11}
	g[l,k]=\dfrac{h^*[L-1-l,-k]e^{-j2\pi k(L-1-l)}}{\sqrt{\sum\limits_{k'=-K/2}^{K/2}\sum\limits_{l'=0}^{L-1}|h[l',k']|^2}},
\end{align}
where $l=0,1,\cdot\cdot\cdot,L-1, k=-K/2,\cdot\cdot\cdot,0,\cdot\cdot\cdot,K/2$. The output signal $\mathbf{\hat{y}} $ can be expressed as
\begin{equation}\label{12}
	\begin{split}
		\begin{aligned}	
		\hat{y}[c]  =\sum\limits_{k = -K/2}^{K/2}\sum\limits_{l=0}^{L-1} g[l,k] y[c-l]_{MN} e^{j2\pi k (c-l)}+ v[c] .
		\end{aligned}
	\end{split}
\end{equation}

We can observe that the output signal $\mathbf{\hat{y}} $ is obtained by the transmitted signal $\mathbf{x} $ sequentially goes through the cascaded DD domain channel $h[l,k]$ and $g[l,k]$. As noted in \cite{b1}, It is equivalent that the signal $\mathbf{x}$ goes through the channel $\hat{h}[l,k]$, where $\hat{h}[l,k]=g[l,k] \star h[l,k]$ is the twisted convolution of $g[l,k]$ and $h[l,k]$. The \eqref{12} can be further expressed as
\begin{equation}\label{35}
 	\begin{split}
 		\begin{aligned}	
 				\hat{y}[c] =\sum\limits_{k = -K}^{K}\sum\limits_{l=0}^{2L-2} \hat{h}[l,k] x[c-l]_{MN} e^{j2\pi k (c-l)}+ \hat{v}[c],
 		\end{aligned}
 	\end{split}
\end{equation}
where $\hat{v}[c]$ is the noise filtered by the channel $g[l,k]$. The result of twisted convolution channel $\hat{h}[l,k]$ can be expressed as
\begin{equation}\label{13}
	\begin{aligned}
		 \hat{h}[l,k] =\frac{	\sum\limits_{k'=-K/2}^{K/2}\sum\limits_{l'=0}^{L-1} h^{*}[l'-\tilde{l},k'-k]h[l',k']e^{j2\pi\tilde{l} (k-k')}}{\sqrt{\sum\limits_{k'=-K/2}^{K/2}\sum\limits_{l'=0}^{L-1}|h[l',k']|^2}},	
	\end{aligned}
\end{equation}
where $ \tilde{l} = l-L+1 $. Through analysis of \eqref{13}, we can observe that the maximum-power
central peak can be achieved as below when $l=L-1, k=0$,

\begin{align}\label{14}
	\hat{h}[L-1,0] =\sqrt{\sum_{k'=-K/2}^{K/2}\sum_{l'=0}^{L-1}|h[l',k']|^2}.
\end{align}

From the \eqref{14}, we observe that the DDR technology can collect the signal energy of all the paths in DD domain channels. 

Considering a special case where all significant paths go through with a small angular spread. The Doppler shifts of all paths in DD domain can be approximately considered equal, then the DD domain channel in \eqref{4} can be reduced to

\begin{align}\label{15}
	h[l,k]=\sum_{l_\tau=0}^{L-1}h_{l_\tau,k_\nu}\delta[l-l_\tau]\delta[k-k_\nu].
\end{align}

Then, \eqref{13} can be reduced to

\begin{equation}\label{17}
	\begin{aligned}
		\hat{h}[l,k_\nu] =\frac{\sum\limits_{l'=0}^{L-1} h^{*}[l'-\tilde{l}, k_\nu]h[l', k_\nu]e^{-j2\pi\tilde{l} k_\nu}}{\sqrt{\sum\limits_{l'=0}^{L-1}|h[l',k_\nu]|^2}}.	
	\end{aligned}
\end{equation}

When there is no Doppler shift (e.g., stationary communication scenario), the $k_\nu=0$, then DD domain channel is simplified to the traditional multipath channel model, \eqref{17} can be further reduced to
\begin{equation}\label{18}
	\begin{aligned}
		\hat{h}[l,0] = \frac{	\sum\limits_{l'=0}^{L-1} h^{*}[l'-\tilde{l} , 0]h[l' , 0]}{\sqrt{\sum\limits_{l'=0}^{L-1}|h[l' , 0]|^2}}.	
	\end{aligned}
\end{equation}

From \eqref{18}, we notice that our proposed DDR technology is equivalent to the classic TR technology in multipath channel model with no Doppler frequency shift \cite{b8}.

To further verify the performance of proposed DDR technology, we compare the gain of SINR between the DDR technology and DP and analyze the computational complexity in the next section.

\section{Performance Analysis}

\subsection{SINR Analysis of DDR}

From \eqref{35}, the output signal $\mathbf{\hat{y}} $  can be further categorized into the desired signal, ISI, inter-Doppler interference (IDI) and noise, as shown below

 \begin{equation}\label{19}	
	\begin{split}	
		\begin{aligned}
		\hat{y}[c] = & \underbrace{\hat{h}[L-1,0]x[c-(L-1)]_{MN}}\limits_{\textnormal{Desired signal}} \\
		& + \underbrace{\underset{l \neq L-1, k \neq 0}{\sum\limits_{k=-K}^{K}\sum\limits_{l=0}^{2L-2}} \hat{h}[l,k] x[c-l]_{MN}e^{j2\pi k (c-l)}}\limits_{\textnormal{ISI, IDI}} \\
		& + \underbrace{\hat{v}[c]}\limits_{\textnormal{Noise}}. \\
		\end{aligned}
	\end{split}
\end{equation}

Then, we can calculate the desired signal power $P_{\textnormal{sig}}^{\textnormal{DDR}}$ as
\begin{equation}\label{20}
	\begin{split}
		\begin{aligned}
			P_{\textnormal{sig}}^{\textnormal{DDR}} &=E_s\left[ \left| \hat{h}[L-1,0]x[c-(L-1)]_{MN}\right| ^2\right] \\
			&=P\left( \left| \hat{h}[L-1,0]\right| ^2 \right) \\
			&=P\left( \sum_{k'=-K/2}^{K/2}\sum_{l'=0}^{L-1}\left| h[l',k']\right| ^2\right), 
		\end{aligned}	
	\end{split}
\end{equation}
where $E_s$ represents the expectation over $\mathbf{x}$. Similarly, the interference power (including ISI and IDI) can be given by
\begin{equation}\label{21}
	\begin{split}
		\begin{aligned}
			&	P_{\textnormal{ISI, IDI}}^{\textnormal{DDR}}  =\\
			& E_s\left[ \left| \underset{l \neq L-1, k \neq 0}{\sum\limits_{k=-K}^{K}\sum\limits_{l=0}^{2L-2}} \hat{h}[l,k] x[c-l]_{MN}e^{j2\pi k (c-l)}  \right| ^2\right] \\	
			&=P\left( \underset{l \neq L-1, k \neq 0}{\sum\limits_{k=-K}^{K}\sum\limits_{l=0}^{2L-2}} \left| \hat{h}[l,k] \right|^2 \right). 
		\end{aligned}	
	\end{split}
\end{equation}

Finally, the $\textnormal{SINR}^{\textnormal{DDR}}$ can be expressed as
\begin{align}\label{22}
	\textnormal{SINR}^{\textnormal{DDR}}=\frac{P_{\textnormal{sig}}^{\textnormal{DDR}}}{P_{\textnormal{ISI, IDI}}^{\textnormal{DDR}}+\sigma^2}.
\end{align}

\subsection{SINR Analysis of DP}

The received signal of DP without DDR can be expressed in \eqref{5}. Similar to the analysis of DDR, we can separate the received signal $\mathbf{y}$ into the desired signal, ISI, IDI and noise, as shown in the following

\begin{equation}\label{23}
	\begin{split}
		\begin{aligned}
			y[c] &= \underbrace{h[\dot{l},\dot{k}] x[c-\dot{l}]_{MN}e^{j2\pi \dot{k} (c-\dot{l})}}\limits_{\textnormal{Desired signal}} \\
			&+ \underbrace{\underset{l \neq \dot{l}, k \neq \dot{k}}{\sum\limits_{k=-K/2}^{K/2}\sum\limits_{l=0}^{L-1}}h[l,k] x[c-l]_{MN}e^{j2\pi k (c-l)}}\limits_{\textnormal{ISI, IDI}} \\
			&+ \underbrace{v[c]}\limits_{\textnormal{Noise}}, \\
		\end{aligned}	
	\end{split} 
\end{equation}
where
\begin{align}\label{24}
	h[\dot{l},\dot{k}]=\underset{l = 0,...,L-1 \atop k = -K/2,...,K/2}{\textnormal{max}} h{[l,k]}.
\end{align}

Then, we can calculate the desired signal power $P_{\textnormal{sig}}^{\textnormal{DP}}$ as
\begin{equation}\label{25}
	\begin{aligned}
		P_{\textnormal{sig}}^{\textnormal{DP}} &=E_s \left[ \left| h[\dot{l},\dot{k}] x[c-\dot{l}]_{MN}e^{j2\pi \dot{k} (c-\dot{l})} \right| ^2  \right] \\
		&=P\left(  \left| h[\dot{l},\dot{k}] \right| ^2 \right).
	\end{aligned}	
\end{equation}

Similarly, the interference power of DP can be given by
\begin{equation}\label{26}
	\begin{aligned}
		P_{\textnormal{ISI, IDI}}^{\textnormal{DP}}&=E_s \left[ \left|\underset{l \neq \dot{l}, k \neq \dot{k}}{\sum\limits_{k=-K/2}^{K/2}\sum\limits_{l=0}^{L-1}}h[l,k] x[c-l]_{MN}e^{j2\pi k (c-l)} \right| ^2  \right] \\
		&=P\left(  \underset{l \neq \dot{l}, k \neq \dot{k}}{\sum\limits_{k=-K/2}^{K/2}\sum\limits_{l=0}^{L-1}} \left|  h[l,k] \right| ^2 \right).
	\end{aligned}	
\end{equation}

Finally, the $\textnormal{SINR}^{\textnormal{DP}}$ can be expressed as
\begin{equation}\label{27}
	\begin{aligned}
		\textnormal{SINR}^{\textnormal{DP}}=\frac{P_{\textnormal{sig}}^{\textnormal{DP}}}{P_{\textnormal{ISI, IDI}}^{\textnormal{DP}} + \sigma^2}.
	\end{aligned}	
\end{equation}

In order to further analyze the performance advantage of DDR technology, it is necessary to analyze the SINR gain between DDR and DP, which is given by

\begin{equation}\label{28}
	\begin{aligned}
		G_p&=\frac{\textnormal{SINR}^{\textnormal{DDR}}}{\textnormal{SINR}^{\textnormal{DP}}} 
		= \frac{P_{\textnormal{sig}}^{\textnormal{DDR}}}{P_{\textnormal{sig}}^{\textnormal{DP}}} \cdot \frac{P_{\textnormal{ISI, IDI}}^{\textnormal{DP}} + \sigma^2}{P_{\textnormal{ISI, IDI}}^{\textnormal{DDR}}+\sigma^2},
	\end{aligned}	
\end{equation} 
where
\begin{equation}\label{29}
	\begin{aligned}
		\frac{P_{\textnormal{sig}}^{\textnormal{DDR}}}{P_{\textnormal{sig}}^{\textnormal{DP}}} & = \frac{P\left( \sum\limits_{k'=-K/2}^{K/2}\sum\limits_{l'=0}^{L-1}\left| h[l',k'] \right| ^2 \right) }{P\left(  \left| h[\dot{l},\dot{k}] \right| ^2 \right)} \\
		& = 1+\frac{ \underset{l' \neq \dot{l}, k' \neq \dot{k}}{\sum\limits_{k'=-K/2}^{K/2}\sum\limits_{l'=0}^{L-1}} \left|  h[l',k'] \right| ^2}{ \left| h[\dot{l},\dot{k}] \right|^2 } > 1.
	\end{aligned}	
\end{equation} 

As most amplitude of paths have been compressed to very low that can be ignored after DDR processing, the interference of ISI and IDI can be approximately considered to depend on few significant paths. Thus, the interference-plus-noise part can be approximately regarded as equal for DDR and DP, which can be written as
\begin{equation}\label{30}
	\begin{aligned}
		\frac{P_{\textnormal{ISI, IDI}}^{\textnormal{DP}} + \sigma^2}{P_{\textnormal{ISI, IDI}}^{\textnormal{DDR}}+\sigma^2} \approx 1 .
	\end{aligned}	
\end{equation} 

Finally, we can observe that the SINR gain $G_p>1$ and there is a significant SINR improvement compared to the DP. The main reason for SINR improvement is that the DDR technology can collect the signal energy of all the paths in DD domain channel, then it can effectively improve the receiver performance of OTFS system in doubly-selective fading channels.

\subsection{Complexity Analysis of DDR}

From the DDR algorithm discussion, we can observe the complexity of the proposed receivers can be attributed to the \eqref{12}, where the number of complex multiplications (CMs) is $\mathcal{O}\left( MNLK \right) $. Considering the channel sparsity in DD domain, there are very few non-zero element in DD domain channel. We assume the number of non-zero element in DD domain channel is $S$, so there are also $S$ non-zero element in DDR channel. The number of CMs required in steps \eqref{12} can further reduce to $\mathcal{O}\left( MNS \right) $. We can observe that there is a lower computational complexity compared to other detection algorithms for OTFS system, such as LMMSE \cite{b6}, MP \cite{b2}, AMP \cite{b3} and UAMP \cite{b5}, whose computational complexities are $\mathcal{O}\left( IM^3 N^3 \right) $, $\mathcal{O}\left( 2^Q IMNS \right) $, $\mathcal{O}\left( 2^Q IMNS \right) $ and $\mathcal{O}\left( IMN \text{log}\left( MN\right)  \right) +  \mathcal{O} \left( 2^Q IMN \right) $, respectively, where $Q$ and $I$ represent the size of modulation alphabet and the number of iterative, respectively.

\section{Simulation Results}
In this section, we test the BER performance of DDR technology in different scenarios, including different number of antennas, modulation alphabet and UE speed. Unless otherwise mentioned, we choose the simulation parameters given in TABLE \ref{tab1}. Firstly, the ideal channel estimation is assumed, i.e, the channel impulse function $h[l,k]$ is perfectly known at the receiver. For OTFS system, extended vehicular model \cite{b10} is adopted to generate the delay taps and each delay tap has a single Doppler shift generated by using Jakes’ formula, i.e., $\nu_i=\nu_{max} cos(\theta_i)$, where $\nu_{max}$ denotes the maximum Doppler shift and is determined by the moving speed. $\nu_i$ represents the Doppler spread of $i$-th path and $\theta_i$ is uniformly distributed over $[-\pi,\pi]$.

\begin{table}[htbp]
	\caption{Simulation Parameters}
	\begin{center}
		\begin{tabular}{|c|c|}
			\hline
			& \\[-7pt]
			Parameter & Value \\
			\hline 
			& \\[-7pt]
			Carrier frequency & 4GHz \\
			\hline 
			& \\[-7pt]
			No. of subcarriers (M) & 512 \\
			\hline 
			& \\[-7pt]
			No. of OTFS symbols (N) & 128 \\
			\hline
			& \\[-7pt]
			Subcarrier spacing & 15KHz \\
			\hline
			& \\[-7pt]
			Modulation alphabet & BPSK, QPSK, 8PSK \\
			\hline
			& \\[-7pt]
			UE speed (Kmph) & 100Kmph, 200Kmph, 300Kmph \\
			\hline
			& \\[-7pt]
			Channel estimation & perfect and imperfect  \\
	 		\hline
	   		& \\[-7pt]
			No. of receiver antennas (Q) & 1,2,4  \\
			\hline	
		\end{tabular}
		\label{tab1}
	\end{center}
\end{table}

\begin{figure}[htbp]
	\centerline{\includegraphics[scale=0.53]{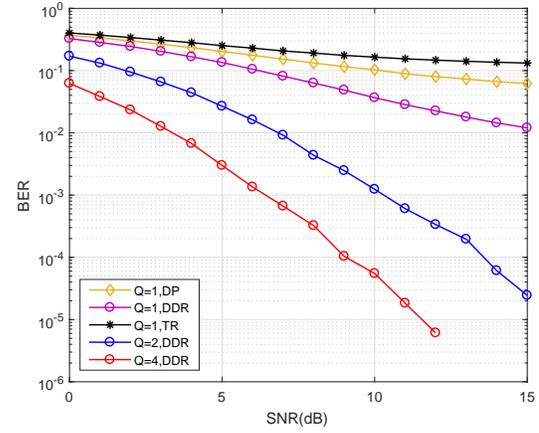}}
	\caption{BER performance comparison for different OTFS receivers.}
	\label{fig1}	
\end{figure}

In Fig. \ref{fig1}, we first compare the BER performance of DDR with perfect time synchronization, DP and classic TR with perfect time-frequency synchronization under OTFS based on QPSK modulation. We observe that the BER performance of DDR is better than DP and classic TR. The main reason for BER performance improvement is coming from the SINR gain of DDR for OTFS systems. The DDR technology can
effectively collect signal energy of all the paths in DD domain channel to improve the receiver performance with low computational complexity $\mathcal{O}\left( MNS \right) $. However, there is an abnormal BER performance between DP and classic TR. In most of existing literatures, the BER performance of classic TR has been proven better than DP in ISI channel. However, the classic TR does not showing any advantages than DP for doubly-selective fading channels. This is due to the fact that Doppler frequency shift will cause severe interference when the classic TR applied directly. In other words, the classic TR only realize the time domain focus and ignore the influence of Doppler shift, leading to significant performance loss. In addition, we observe that better performance can be achieved as the number of antennas increases due to the additional spatial diversity.


\begin{figure}[htbp]
	\centerline{\includegraphics[scale=0.53]{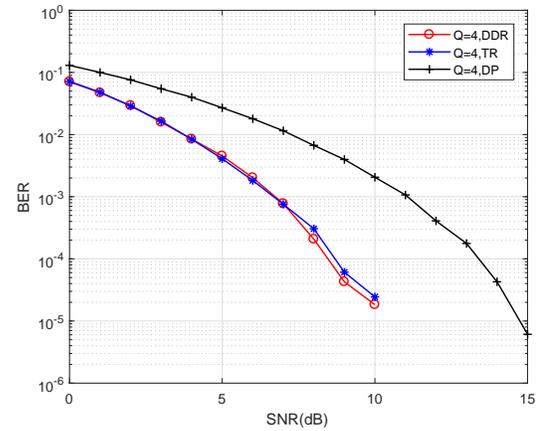}}
	\caption{BER performance comparison of different receivers with channel paths experience similar Doppler spread.}
	\label{fig3}	
\end{figure}

We further consider a special case where each path experience similar Doppler spread (see \eqref{15}) and test the BER performance in Fig. \ref{fig3}. We can observe that the BER performance of classic TR is similar to DDR and both of them are better than DP under such scenario. The main reason is that when the similar Doppler shift is considered for all the paths, the classic TR with perfect time-frequency synchronization can potentially address the interference caused by the Doppler shift, resulting in a similar BER performance as that of DDR.

\begin{figure}[htbp]
	\centerline{\includegraphics[scale=0.53]{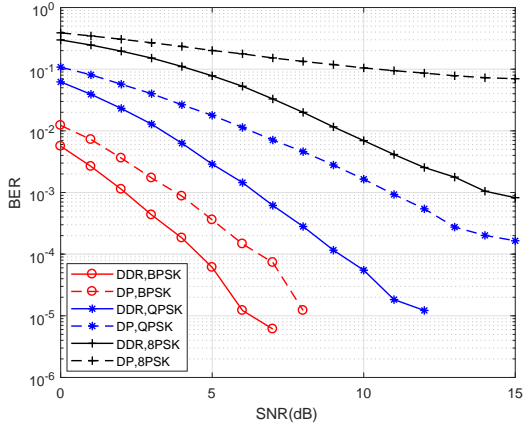}}
	\caption{BER performance comparison of OTFS with BPSK,QPSK and 8PSK.}
	\label{fig4}	
\end{figure}

\begin{figure}[htbp]
	\centerline{\includegraphics[scale=0.53]{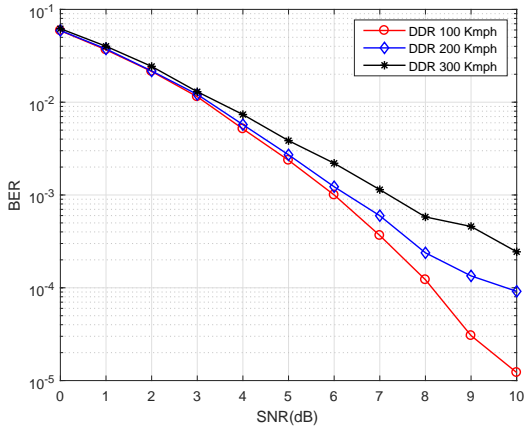}}
	\caption{BER performance comparison of DDR receiver for different UE speed.}
	\label{fig5}	
\end{figure}

Fig. \ref{fig4} shows the BER performance of OTFS with BPSK, QPSK and 8PSK when $ Q=4 $. The results is very obvious, the high order modulation alphabet achieves higher spectral efficiency but worse BER performance compared to low order modulation alphabet scenario. Fig. \ref{fig5} illustrates the BER performance of DDR receiver for different UE speed when $Q=4$. We can notice that even the BER performance is slightly decreases as the UE speed increases, our proposed DDR receiver still function for high mobility communications.

\begin{figure}[htbp]
	\centerline{\includegraphics[scale=0.53]{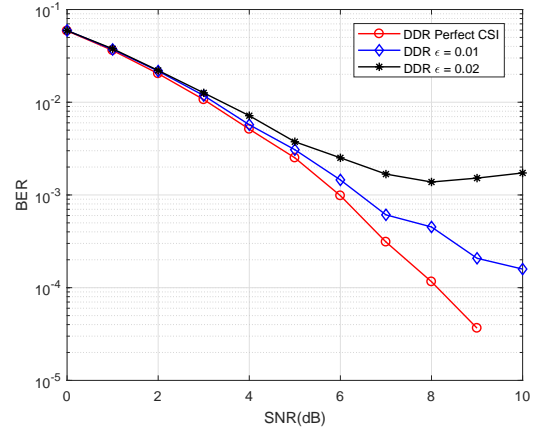}}
	\caption{BER performance of DDR receiver with imperfect CSI.}
	\label{fig6}	
\end{figure}

In previous discussions, we assume the result of channel estimation is perfect, but in practice, the receiver can only acquire CSI based on pilots and training which consume power and spectrum resources. It is therefore common that receivers must function under CSI uncertainty. Here, we characterize the CSI error by adopting the following model\cite{b11}

\begin{equation}\nonumber
	\begin{aligned}
	& h_i=\tilde{h} + \Delta h_i & \Vert\Delta h_i \Vert \leq \epsilon_{h_i}, \\
	& \tau_i=\tilde{\tau} + \Delta \tau_i & \Vert\Delta \tau_i \Vert \leq \epsilon_{\tau_i}, \\
	& \nu_i=\tilde{\nu} + \Delta \nu_i & \Vert\Delta \nu_i \Vert \leq \epsilon_{\nu_i}, 
	\end{aligned}	
\end{equation} 
where $\tilde{h}$, $\tilde{\tau}$ and $\tilde{\nu}$ are the estimated versions of $h_i$, $\tau_i$ and $\nu_i$. $\Delta h_i$, $\Delta \tau_i$ and $\Delta \nu_i $ represent the corresponding channel estimation errors, whose norms are bounded with the given radius $\epsilon_{h_i}$ , $\epsilon_{\tau_i}$ and $\epsilon_{\nu_i}$. For simplicity, we assume that $\epsilon_{h_i} =  \epsilon \Vert h_i \Vert $ , $\epsilon_{\tau_i} =  \epsilon \Vert \tau_i \Vert $ and $\epsilon_{\nu_i} =  \epsilon \Vert \nu_i \Vert $. From Fig. \ref{fig6}, we can observe that DDR technology can tolerate a certain degree of channel uncertainty $\epsilon$. Without sudden and large drop of receiver performance as channel uncertainty grows, our proposed DDR receiver is robust and can handle typical CSI errors.

\section{Conclusion}

In this paper, we developed the DDR technology from a perspective of two-dimensional cascaded channel model for OTFS system and verified that there is a significant SINR improvement compared to the DP with low complexity by both theoretical analysis and numerical test. Through the sufficient simulations, we evaluated the BER performance in different scenarios and schemes, including the different number of antennas, modulation alphabet and UE speed. Finally, we conclude that our proposed DDR technology can effectively improve the performance of OTFS system compared to traditional DP and TR receivers with low complexity in doubly-selective fading channels, and also robust to the imperfect CSI.

\end{document}